\documentclass[12pt,preprint]{aastex}

\shorttitle{Properties of intranetwork horizontal magnetic
elements} \shortauthors{Jin et al.}

\begin{document}


\title{THE PROPERTIES OF HORIZONTAL MAGNETIC ELEMENTS IN QUIET SOLAR INTRANETWORK}

\author{Chunlan Jin, Jingxiu Wang and Guiping Zhou}

\altaffiltext{}{National Astronomical Observatories, Chinese
Academy of Sciences, Beijing 100012, China; E-mail:
jinchunlan@ourstar.bao.ac.cn, wangjx@ourstar.bao.ac.cn,
gpzhou@ourstar.bao.ac.cn}

\begin{abstract}
Using the data observed by the Solar Optical
Telescope/Spectro-Polarimeter aboard the \emph{Hinode} satellite,
the horizontal and vertical fields are derived from the
wavelength-integrated measures of Zeeman-induced linear and
circular polarizations. The quiet intranetwork regions are
pervaded by horizontal magnetic elements. We categorize the
horizontal intranetwork magnetic elements into two types: one is
the non-isolated element which is accompanied by the vertical
magnetic elements during its evolution; another is the isolated
element which is not accompanied by the vertical magnetic
elements. Their properties, such as lifetime, size and magnetic
flux density, are studied, and the relationships among various
magnetic parameters are investigated.

We identify 446 horizontal intranetwork magnetic elements, among
them 87 elements are isolated and 359 are non-isolated. Quantitative
measurements reveal that the isolated elements have relatively
weaker horizontal magnetic fields, almost equal size, and shorter
lifetime comparing with the non-isolated elements. Most non-isolated
horizontal intranetwork magnetic elements are identified to
associate with the emergence of $\Omega$-shaped flux loops. A few
non-isolated elements seem to indicate scenarios of submergence of
$\Omega$ loops or emergence of U-like loops. There is a positive
correlation between the lifetime and the size for both the isolated
and non-isolated HIFs. It is also found that there is also positive
correlation between the lifetime and the magnetic flux density for
non-isolated HIFs, but no correlation for isolated HIFs. Even though
the horizontal elements show lower magnetic flux density, they could
carry the total magnetic flux in the order of magnitude close to
$10^{25}$ Mx to the solar surface each day.

\end{abstract}

\keywords{Sun: evolution --- Sun: magnetic fields --- Sun:
photosphere --- techniques: polarimetric}

\section{INTRODUCTION}

It has been known for more than 30 years that the quiet
photosphere contains small-scale intranetwork (inter-network,
inner-network) (IN) magnetic fields (Smithson 1975; Livingston \&
Harvey 1975). Most IN magnetic observations have been made at or
near disk center using circular-polarization signals that reveal
properties of the vertical component of IN fields, such as the
intrinsic field strength (e.g., Keller et al. 1994; Lin 1995;
L\'{o}pez Ariste 2006; S\'{a}nchez Almeida 2006 ), internal
structures (e.g., S\'{a}nchez Almeida et al. 1996; Lites \&
Socas-Navarro 2004), and so on. The most prominent are the flux
elements of the order of $10^{16}$ Mx (Wang et al. 1985; Wang et
al. 1995; Lin \& Rimmele 1999) and their rapid evolution (Shi et
al. 1990; Zhang et al. 1998; Berger et al. 1998; Dom\'{i}nguez
Cerde\~{n}a et al. 2003). The small fluxes and sizes with rapid
time changes make the IN magnetic fields provide important
contribution to the Sun's magnetism (Wang et al. 1995; Meunier et
al. 1998; S\'{a}nchez Almeida \& Lites 2000; Lites 2002;
S\'{a}nchez Almeida 2003; Khomenko et al. 2005; Dom\'{i}nguez
Cerde\~{n}a et al. 2006a). Moreover, the IN magnetic fields also
have important effect on heating solar chromosphere and corona
(e.g., Zhang \& Zhang 2000; S\'{a}nches Almeida et al. 2004).

However, there is scant information of the IN horizontal magnetic
component. By the distinct signature of Stokes $Q$ and $U$
polarization profiles in quiet regions close to the Sun's disk
center based on the observation with the Advanced Stokes
Polarimeter, Lites et al. (1996) firstly revealed the isolated,
small-scale (typically 1$''$ - 2$''$ or smaller), short-lived
horizontal IN field elements that were abbreviated as HIFs. Meunier
et al. (1998) made one-dimension scans across the disk and concluded
that the IN magnetic field consisted of relatively stronger vertical
features and weaker horizontal component. Harvey et al. (2007)
discovered the ubiquitous horizontal component of the solar magnetic
field in quiet regions of the photosphere using the observations
with different instruments, spectrum lines, and measurements
techniques. They found that the horizontal component exhibited wide
ranges of spacial and temporal scales: from a few arcseconds up to
15 arcseconds and from several minutes to hours, respectively. The
observations from the high spacial resolution of the Solar Optical
Telescope (SOT: Tsuneta et al. 2008a; Suematsu et al. 2008; Ichimoto
et al. 2008; Shimizu et al. 2008) aboard \emph{Hinode} (Kosugi et
al. 2007) and the good polarimetric precision of the
Spectro-Polarimeter (SP: Tarbell et al. 2007) provide an access to a
new regime of the product of angular resolution and polarimetric
precision, and a few papers are concerned with the IN horizontal
field. Centeno et al. (2007) study the emergence of magnetic flux at
very small spatial scales in the quiet-Sun IN. They find that the
horizontal magnetic field appears prior to any significant amount of
vertical field. As time goes on, the traces of the horizontal field
disappear, while the vertical dipoles are carried by the plasma
motions toward the surrounding intergranular lanes. Orozco
Su\'{a}rez et al. (2007a, 2007b) show that IN consists of very
inclined, hG fields. Ishikawa et al. (2008) reveal widespread
occurrence of transient, spatially isolated horizontal magnetic
fields, and the lateral extent of the horizontal magnetic fields is
comparable to the size of the photospheric granules. Lites et al.
(2008) reveal that the quiet IN regions are pervaded by horizontal
magnetic flux; the horizontal magnetic flux is not associated with
network flux. They declare that horizontal magnetic flux is an IN
phenomenon. The horizontal fields are somewhat separated spatially
from the vertical fields. On average, the horizontal magnetic flux
density is apparently larger than the vertical one. Ishikawa \&
Tsuneta (2009) compared the properties of the transient horizontal
magnetic fields in both plage and quiet Sun regions. They show that
there are no differences in the distribution and occurrence rate of
horizontal fields between the quiet Sun and the plage regions, and
there is essentially no preferred orientation for transient
horizontal magnetic fields in either region.

In a word, the advance in spatial resolution and sensitivity of
SOT/SP observations afforded by \emph{Hinode} presents us with an
opportunity to explore the horizontal fields in the quiet IN region,
and a new era of studies for horizontal magnetic fields on the quiet
Sun is opening.

Here, by the time sequence of solar photospheric observations near
the disk center using SOT/SP aboard the \emph{Hinode} satellite,
we identify and analyze 87 isolated HIF elements, and 359
non-isolated elements which are 4 times more numerous than that of
isolated ones. Their properties, such as the size, lifetime, and
magnetic flux density are studied. Furthermore, the relationships
among the parameters representing various magnetic properties are
investigated. The global contribution of HIFs to the magnetic flux
on solar surface is estimated.

In Section 2, we show the observations from SOT/SP aboard the
\emph{Hinode} satellite for the selected regions. In Section 3, we
describe the technique used to extract horizontal and vertical
magnetic fields from wavelength-integrated measures of
Zeeman-induced linear polarization and circular polarization. In
section 4, we analyze the size, lifetime, and horizontal magnetic
flux density, study the relationships among the magnetic
parameters, and give the estimation of the contribution of HIFs to
the Sun's magnetic flux. In Section 5, we discuss the topology
configuration. In Section 6, we give the summary.

\section{OBSERVATIONS}

The SP observations in the SOT instruments aboard \emph{Hinode}
spacecraft provide the full Stokes spectral signals of two
magnetically sensitive FeI lines at 630.15 nm ($g_{eff}=1.67$) and
630.25 nm ($g_{eff}=2.5$). The spatial resolution of SP
observation is 0$''$.32, and the wavelength sampling is 2.16 pm.

We select 15 time-sequences of fast mode observations close to the
disk center, which include 11 quiet regions, 2 ephemeral regions and
2 plage regions ( see Table 1). The data in each time-sequence
consist of a series of maps with the field of view (FOV) 8$''$.86
$\times$ 162$''$.30. Each map is composed of 30 consecutive
positions of spectrograph slit, and the temporal resolution for each
map is $\sim$ 2 minutes. In addition, for estimating the magnetic
flux carried to the solar surface by the HIFs, we select a quiet
solar region of fast mode observation with FOV 118$''$.89 $\times$
162$''$.30, which was observed at disk center on 2007 May 11 (see
Table 1). The region consists of 403 consecutive positions of
spectrograph slit.

The scanning steps of the spectrograph slit of fast mode observation
are 0$''$.295. For the observation of each spectrograph slit, the
exposure time is 3.2 seconds. The polarization in solar continuous
spectrum of wavelength 630 nm is few according to the results
obtained by Stenflo (2005). Therefore, we consider the polarization
of continuous spectrum 630 nm from the observation of SP as noise,
and derive a noise level of $1.16\times 10^{-3}I_{c}$ in Stokes $Q$
and $U$, and $7.5\times 10^{-4}I_{c}$ in Stokes $V$. Figure 1 shows
a randomly selected example of Stokes spectra from the near
disk-center time series observation of quiet region.

\section{CONVERSION OF POLARIZATION SIGNALS TO MAGNETIC FLUX DENSITY}

In this paper, we mostly focus on HIFs, i.e., the horizontal
magnetic field elements from the IN fields. In fact, the IN fields
have low polarization degree, and mixed and nearly balanced
polarity in small scale. Thus, the circular and linear
polarization signals from IN fields are very small compared to
those from active regions, and extracting the properties of
magnetic field from IN is confronted with the measure noise.

Although the inversion codes are robust when applied to the strong
Stokes polarization signals from active region, they encounter
difficulties in convergence toward and uniqueness of solutions
when confronted with noisy profiles (Lites et al. 2008).
Therefore, we adopt the procedure of polarization integrated in
wavelength of FeI 630.25 nm line to provide the measure of
longitudinal and transverse fields (Lites et al. 1999). By this
method, we enhance the sensitivity to weak Zeeman polarization in
the presence of measurement noise, and avoid the problems of
convergence and non-uniqueness that result from the inversion of
noisy profiles.

 The wavelength-integrated circular polarization $V_{tot}$ from Stokes $V(\lambda)$ is defined as

\begin{equation}
V_{tot}=sign(V_{blue})\frac{|\int_{\lambda_{b}}^{\lambda_{0}}V(\lambda)d\lambda|+|\int_{\lambda_{0}}^{\lambda_{r}}V(\lambda)d\lambda|}{I_{c}\int_{\lambda_{b}}^{\lambda_{r}}d\lambda}
\end{equation}
where sign($V_{blue}$) means the sign of the blue peak of the
Stokes $V$ profile, $\lambda_{0}$ describes the wavelength of line
center (i.e., the wavelength corresponding to the minimum value of
$I(\lambda$)), $\lambda_{r}$ and $\lambda_{b}$ show the integrated
limit of the red and the blue over the line FeI 630.25 nm, with
$\lambda_{r}$=$\lambda_{0}$+35 pm and
$\lambda_{b}$=$\lambda_{0}$-35 pm, $I_{c}$ means the continuum
intensity. Furthermore, we also define the linear polarization
$L_{tot}$ with the same method,

\begin{equation}
L_{tot}=\frac{\int_{\lambda_{b}}^{\lambda_{r}}[Q^{2}(\lambda)+U^{2}(\lambda)]^{1/2}d\lambda}{I_{c}\int_{\lambda_{b}}^{\lambda_{r}}d\lambda}
\end{equation}
Where $Q(\lambda)$ and $U(\lambda)$ is the intensity of Stokes $Q$
and $U$ in wavelength $\lambda$.

 Jefferies, Lites, \& Skumanich (1989)
demonstrated that, in the weak field limit

\begin{equation}
fB^{2}sin^{2}\gamma \propto
(\int_{0}^{\infty}[Q^{2}(\lambda)+U^{2}(\lambda)]^{1/2}d\lambda)/D
\end{equation}

\begin{equation}
fBcos\gamma \propto (\int_{0}^{\infty}|V(\lambda)|d\lambda)/D
\end{equation}
Where $f$ is the filling factor, $B$ is the magnetic field
strength, $\gamma$ is the inclination angle corresponding to the
line of sight (LOS), and $D=I_{c}-I(\lambda=0)$ is the line depth
at line center. Therefore, Lites at al. (1999) used inversions of
the stronger polarization signals in quiet solar
spectropolarimetric maps to calibrate circular polarization into
longitudinal fields. Following that work, we inverted the pixels
with stronger linear and circular polarization signals primarily
from the network flux elements to determine the calibration
constants relating linear polarization to transverse fields and
circular polarization to longitudinal fields.

Mart\'{i}nez Gonz\'{a}lez et al. (2006) demonstrate the pair of FeI
lines in low magnetic flux quiet Sun regions is not capable of
distinguishing between the intrinsic magnetic field and the filling
factor. Therefore, the magnetic flux density is a more appropriate
quantity to describe the equivalent, spatially resolved vector
magnetic fields, i.e., the longitudinal and the transverse
components, $B^{L}_{app}$ and $B^{T}_{app}$. The longitudinal
component $B^{L}_{app}$ may be thought of as the magnitude of the
LOS component of a spatially resolved magnetic field that produces
the observed circular polarization signal, while the transverse
component $B^{T}_{app}$ is perpendicular to the LOS that would
produce the observed linear polarization signal. Figure 2 shows the
relationships of measured $V_{tot}$ as a function of the
longitudinal flux density $B^{L}_{app}$ inferred from the inversion,
$f|B|$cos($\gamma$), and measured $L_{tot}$ as a function of the
transverse flux density $B^{T}_{app}$ derived from the inversion,
$f^{1/2}|B|$sin($\gamma$). We have used the inversion code developed
by the Japanese \emph{Hinode} group based on the assumption of
Milne-Eddington atmosphere (T. Yokoyama 2009, in preparation).

\section{PROPERTIES OF HORIZONTAL INTRANETWORK MAGNETIC ELEMENTS}

\subsection{Size, Lifetime, And Horizontal Magnetic Flux Density}

For each horizontal magnetic element, we begin to track it when its
linear polarization signal is obviously larger than the noise level,
and we think it disappears when its linear polarization is difficult
to distinguish from the noise. We selected 446 independent HIFs by
identifying and tracking each of them from birth to death. They
cover almost all the prominent HIFs seen from the 15 time-sequence
of the quiet sun horizontal field measurements. We quantitatively
measure the lifetime, size and average horizontal magnetic flux
density, respectively, for each horizontal magnetic element.

Lites et al. (2008) point out that the horizontal fields do not
simply arise from convective buffeting of strong, nearly vertical
fields. Symmetric Stokes $Q$ and $U$ signatures are not accompanying
strong antisymmetric Stokes $V$ profiles (Lites et al. 1996).
Moreover, only 16\% of the area within the 3$\sigma$ horizontal
fields contours overlaps with area within the contours of vertical
fields (Lites et al. 2008). Therefore, we categorized the HIFs into
two classes. If the circular polarization signals tempo-spatially
accompanying the HIFs are always less than the noise level of
circular polarization during the evolution of HIFs, then we would
define these horizontal elements as isolated HIFs. To the contrary,
if there are clear circular polarization signals above the noise
tempo-spatially accompanying the HIFs, we would called them as
non-isolated HIFs. Thus, it is found that there are 87 isolated HIFs
and 359 non-isolated HIFs. A case of evolution non-isolated HIFs is
displayed in the top two panels of Fig. 3. From this figure, we can
find that there are always two vertical magnetic elements with
opposite polarities on both sides of the HIF (denoted by dotted
lines) during its evolution, and the two vertical magnetic elements
still exist after disappearance of the HIF. Moreover, the two
vertical magnetic elements (denoted by arrows at 07:56 UT) with
opposite polarities separate. An example of isolated HIFs (displayed
by dotted lines) is shown in the lower two panels of Fig. 3. For the
isolated HIF, there are no vertical magnetic elements during its
evolution, even after its disappearance.

Because the spatial size and average horizontal magnetic flux
density are changing during the process of HIFs' evolution, we
define the largest size during the evolution as the spatial size of
HIFs and the largest horizontal magnetic flux density during the
evolution as the horizontal magnetic flux density of HIFs.
Furthermore, we define the lifetime of HIF according to the number
of maps of the appearance of HIF. That is to say, in the maps of
time series, if there is only one map which shows the HIF, we will
define the lifetime of the HIF as 2 minutes, though it may not be
exactly true. The properties of HIFs are shown in Table 2.

Figure 4 shows the histogram of the lifetime of HIFs. The dotted
line shows the histogram of lifetime for isolated HIFs, the
dash-dot-dot line for non-isolated HIFs, and the solid line for
all HIFs. From the figure, we can find that the lifetime of these
isolated HIFs ranges from 2 minutes to 14 minutes with a peak
distribution at 4 minutes, while the lifetimes of these
non-isolated HIFs range from 2 minutes to 14 minutes with a peak
distribution at 6 minutes. The mean lifetime is 4.3 minutes for
isolated HIFs and 5.9 minutes for non-isolated HIFs, respectively.
On average, the lifetime of isolated HIFs is shorter than that of
non-isolated HIFs.

Figure 5 shows the histograms of the size of all HIFs (displayed
by the solid line), the isolated HIFs (showed by dotted line) and
the non-isolated HIFs (denoted by dash-dot-dot line),
respectively. Because the shapes of these horizontal magnetic
elements are irregular, the size of horizontal magnetic element is
defined as follows: Let A be the area of horizontal magnetic
element, and an effective size is $D=2(A/\pi)^{1/2}$. The size of
these isolated HIFs ranges from 0$''$.62 to 1$''$.48 with a peak
distribution at about 1$''$.00, while the size of these
non-isolated HIFs ranges from 0$''$.72 to 2$''$.04 with a peak
distribution at about 1$''$.15. The mean size is 1$''$.07 for
isolated HIFs and 1$''$.14 for non-isolated HIFs. On average, the
size of isolated HIFs is less than that of non-isolated HIFs.

Figure 6 shows the histogram of the horizontal magnetic flux
density of HIFs. The horizontal magnetic flux density of these
isolated HIFs range from 247 Mx cm$^{-2}$ to 304 Mx cm$^{-2}$ with
a peak distribution around 257 Mx cm$^{-2}$, while the horizontal
magnetic flux density of these non-isolated HIFs range from 248 Mx
cm$^{-2}$ to 330 Mx cm$^{-2}$ with a peak distribution at about
267 Mx cm$^{-2}$. The mean horizontal magnetic flux density is 264
Mx cm$^{-2}$ for isolated HIFs and 273 Mx cm$^{-2}$ for
non-isolated HIFs. The horizontal magnetic flux density of
isolated HIFs is averagely smaller than that of non-isolated HIFs.

In order to test whether the difference of properties between
isolated HIFs and non-isolated HIFs is remarkable, we make the
statistical test. Here, we give the detail description of the
statistical test. Taking an example of the difference between the
average sizes of isolated HIFs and non-isolated HIFs. The steps are
following: (1) we assume $H_{0}$: the average size of isolated HIFs
is equal to that of non-isolated HIFs; (2) we confirm the
distribution of the statistical variable. In general, when the
number of sample is larger than 30, the sample is considered to a
large sample. For a large sample, a normal distribution for the
average value of the sample can be considered according to the
central limit theorems; (3) a significance level $\alpha$ is given.
Here, the value $\alpha$ is assumed to be 0.1\% by us; (4) the
confidence limit is checked according to the table of normal
distribution.
$\int_{-k_{\frac{\alpha}{2}}}^{k_{\frac{\alpha}{2}}}\frac{1}{\sqrt{2\pi}}e^{-\frac{v^{2}}{2}}dv=1-\alpha$.
We can obtain the $k_{0.0005}=3.29$ by checking the table of normal
distribution; (5) the statistical variable is computed by the
formula:
$u=\frac{\overline{x}_{1}-\overline{x}_{2}}{\sqrt{\frac{\sigma_{1}^{2}}{n_{1}}+\frac{\sigma_{2}^{2}}{n_{2}}}}$,
where $\sigma_{1}^{2}$ is the sample variance of the non-isolated
HIFs, $n_{1}$ is number of non-isolated HIFs, $\sigma_{2}^{2}$ is
the sample variance of the isolated HIFs, and $n_{2}$ is number of
isolated HIFs. Here, $\overline{x}_{1}$ is the average size of
non-isolated HIFs, and $\overline{x}_{2}$ is the average size of
isolated HIFs. We compute that the value of statistical variable is
3.05; (6) the statistical verdict is given: when
$|u|<k_{\frac{a}{2}}$, the assumption of $H_{0}$ is right; when
$|u|\geq k_{\frac{a}{2}}$, the assumption of $H_{0}$ is wrong.
Because the statistical variable $u$ is less than
$k_{\frac{\alpha}{2}}$ by our computation, the assumption of the
same size of isolated HIFs and non-isolated HIFs is right. That is
to say, the average size of isolated HIFs is equal to that of
non-isolated HIFs with a confidence level of 99.9\%.

Following above steps, we also test the distinctions of average
lifetime and average horizontal magnetic flux density between
isolated HIFs and non-isolated HIFs, and we find that the average
lifetime and average horizontal magnetic flux density of isolated
HIFs are less than those of non-isolated HIFs with a confidence
level of 99.9\%. Although the distinction between the isolated and
non-isolated HIFs is clear in a sense of statistics, it is not clear
whether or not the two kinds of HIFs have different origins and if
they represent intrinsically different types of magnetic fields.

\subsection{Relationships Among The Magnetic Parameters}

The correlation between the horizontal magnetic flux density and
lifetime of HIFs is studied. In order to show the relationship
clearly, we divide the horizontal magnetic flux density into 7 bins
for non-isolated HIFs, and 5 bins for isolated HIFs according to the
lifetime (range from 2 minutes to 14 minutes for non-isolated HIFs,
and from 2 minutes to 10 minutes for isolated HIFs). Moreover, we
compute the mean horizontal magnetic flux density in each bin. The
correlation is shown in Fig. 7. From the figure, we find that there
is a positive correlation between them for non-isolated HIFs, and
the correlation coefficient is 0.917, well above the confidence
level of 99.9\%. However, we find that there is no relationship
between them for isolated HIFs due to a small correlation
coefficient of 0.053.

In addition, by the same way, we also analyze the relationship
between the lifetime and the corresponding size of HIFs, just as
shown in Fig. 8. Our statistics display that there is a positive
correlation between them for both non-isolated HIFs and isolated
HIFs, and their correlation coefficients are 0.924 and 0.986,
respectively, both of them well above the confidence level of
99.9\%.

\subsection{Contribution of HIFs to The Sun's Magnetic Flux}

$B^{T}_{app}$ approximately describes the horizontal field strength
if the HIFs were actually completely resolved in our observation.
 Furthermore, there are two methods to obtain
the horizontal magnetic field of each pixel. One is the horizontal
magnetic flux density by allowing for variable filling fraction in
the inversion. Another is the horizontal field strength by assuming
the resolved HIFs in the observation and unit filling fraction in
the inversion. The horizontal magnetic fields obtained by the two
methods yield very nearly the same average value (Lites et al.
1996).

Considering the evolution of HIFs, we estimate the magnetic flux
carried to the solar surface by these horizontal elements by
assuming the variable filling fraction in the inversion. The
magnetic flux in one HIFs, $\psi\approx B^{T}_{app}Dd$ is about
$2.2\times 10^{17}$ Mx for horizontal magnetic flux density
$B^{T}_{app}$ of 271 Mx cm$^{-2}$, the size $D$ of 1$''$.13 (820 km)
and assuming a vertical extension in the atmosphere of 100 km
(comparable to the photospheric scale height, so that the flux
elements are actually observable). In figure 9, we find that the
quiet IN regions are pervaded by horizontal magnetic elements, and
the mean number density of HIFs is $\sim$ 0.015 per arcsec$^{2}$.
Given that the mean lifetime of HIFs is $\sim$ 5.6 minutes, we can
obtain the rate of $4.46\times10^{-5}$ HIFs arcsec$^{-2}$s$^{-1}$.
Therefore, almost $9.82\times10^{24}$ Mx over the solar surface per
day, or $3.94\times10^{28}$ Mx per 11 year sunspot cycle is carried
to solar surface by the HIFs. The flux emergence rate is similar to
the estimation of the HIFs by Lites et al. (1996) and the IN
vertical flux emergence by Wang et al. (1995). Moreover, the
horizontal fields also contribute to the "hidden" turbulent flux
suggested by studies involving Hanle effect depolarization of
scattered radiation (Lites et al. 2008). Tsuneta et al. (2008b)
point that the polar region is also covered with ubiquitous
horizontal fields. Thus, even if we exclude 30\% of the solar
surface for network and active region, this is still more than the
accepted value of the total flux though to emerge in bipolar sunspot
regions during an entire solar cycle, of order $10^{25}$ Mx (Harvey
1993). The result agrees with the conclusion drawn by
 Dom\'{i}nguez Cerde\~{n}a (2006b) who find that the quiet Sun
photosphere has far more unsigned magnetic flux than the active
regions and the network together. Indeed, the solar IN fields
contribute the most of the solar magnetic flux in the solar
surface.

\section{DISCUSSION OF THE TOPOLOGY CONFIGURATION}

With regard to the magnetic flux shown in the photosphere, there are
two possible topology configurations for the magnetic loop: one is
the $\Omega$ loop for traditional emerging flux region (Bruzek 1967;
Zwaan 1978), and another is $U$ loop, suggested for IN magnetic
elements by Spruit et al. (1987). For the both types of topology,
the horizontal fields are associated with the vertical fields, and
the footpoints of magnetic loop stand at each side of the region of
linear polarization signals. However, the detailed scenario of the
vertical and horizontal field evolution for the non-isolated HIFs
would tell which topology is truly correct. If the vertical fields
are separating and growing during the HIF's appearance, then we are
quiet sure the HIF represents an emerging $\Omega$ flux loops. On
the other hand, if the HIF appeared during the shrinkage and
reduction of vertical fields, we should expect an emerging $U$ loop.
The submergence of an $\Omega$ would show the same scenario as $U$
loop emergence, but the horizontal field orientation would be
opposite to the later case.

Keeping the above discussion in mind we find the magnetic topology
configurations for the $\Omega$ loop that are shown in the top two
panels of Fig. 10. At 14:59 UT, the horizontal field shows up in
the photosphere as a non-isolated HIF. At 15:07 UT, the horizontal
element disappears while the corresponding vertical magnetic
fields (denoted by circles) with opposite polarities separate. In
our studies, some non-isolated HIFs only display one footpoint of
in a side of horizontal fields. A similar example is shown in the
middle two panels of Fig. 10. From 03:15 UT to 03:23 UT, the
horizontal magnetic element is being accompanied by the positive
vertical magnetic element. This is not conflicting with the
traditional picture of flux emergence. Wang and Shi (1993)
demonstrated that when an emerging flux region appeared in an area
of magnetic flux with one dominant polarity, the subsurface
magnetic reconnection could hide one polarity footpoint shown up
in the photosphere.

However, under the conditions of spatial resolution of 0$''$.32
and temporal resolution of $\sim$ 2 minutes, we can always not
distinguish if the observed magnetic configurations are $\Omega$
loop or $U$ loop, according to the evolutions of horizontal and
vertical magnetic elements. A case is shown in the bottom two
panels of Fig. 10. From 11:54 UT to 11:56 UT, two negative
vertical magnetic elements and a positive vertical magnetic
element are going with the horizontal magnetic element (denoted by
dotted lines). At 11:58 UT, the horizontal magnetic element
suddenly disappears, and the vertical magnetic flux shows mutual
flux disappearance, i.e., flux cancellation (Live et al. 1985;
Martin et al. 1985). During the flux cancellation, a negative
vertical magnetic element almost vanishes. Without resolving the
180 degree ambiguity of field azimuth in the observed vector
fields, we have difficulty to finally determine if the observed
HIF evolution represents a submergence of an $\Omega$ loop or the
emergence of an $\Omega$ loop.

In addition, the distinction in properties between isolated and
non-isolated HIFs seem to offer a rather consistent picture about
the HIFs. The non-isolated HIFs are likely to be the stronger
horizontal field threads which have already overstepped the criteria
of buoyancy instability; thus they emerge in the form of $\Omega$
loops, showing up the non-isolated HIFs. The isolated HIFs come from
weaker horizontal field threads beneath the photosphere. They appear
as the isolated HIFs by pumping force from local convection.
Therefore it is reasonable to propose that beneath the photosphere
there are sea of horizontal magnetic threads which manifest as HIFs
in the photosphere either by buoyancy instability or the pumping
force of the local convection.

\section{SUMMARY}

Using the data observed by the SOT/SP aboard the \emph{Hinode},
the properties of HIFs, such as the size, lifetime, and horizontal
magnetic flux density of HIFs are studied, and the relationships
among these properties of HIFs are investigated.

We select 446 HIFs by identifying and tracking each of them from
birth to death, and find that there are 87 isolated HIFs and 359
non-isolated HIFs which is more numerous than that the former by a
factor of 4. We find that the lifetime of HIFs ranges from 2 minutes
to 14 minutes with a peak lifetime of 6 minutes, and the result is
obviously different from the conclusion drawn by Harvey et al.
(2007) for seething horizontal magnetic fields in the quiet solar
photosphere. A wide range of temporal scales from several minutes to
hours was reported for their seething fields. More studies are
needed to clarify the discrepancy, and particularly if different
authors are dealing with different species of horizontal fields. The
mean lifetime of HIFs is 5.6 minutes, and the result confirms the
conclusion drawn by lites et al. (1996) who find that the horizontal
elements are short-lived, typically lasting $\sim$ 5 minutes. On
average, the lifetime of isolated HIFs is shorted than that of
non-isolated HIFs. The range of size is from 0$''$.62 to 1$''$.48
for isolated HIFs  with a peak distribution at about 1$''$.00, and
from 0$''$.72 to 2$''$.04 for non-isolated HIFs with a peak
distribution at about 1$''$.15. The size of HIFs obtained by us
agrees with the size of horizontal magnetic flux structures
presented by Lites et al. (1996) who firstly reveal the small-scale
(typically 1$''$-2$''$ or smaller) horizontal elements. We find that
the size of isolated HIFs is averagely equal to that of non-isolated
HIFs in a sense of statistics. However, the horizontal magnetic flux
density of isolated HIFs is less than that of non-isolated HIFs on
average.

There is a positive correlation between the lifetime and the size
for both the isolated and non-isolated HIFs. It is also found that
there is also positive correlation between the lifetime and the
magnetic flux density for non-isolated HIFs, but no correlation
for isolated HIFs. Even though the horizontal elements show lower
magnetic flux density, they could carry the total magnetic flux in
the order of magnitude close to $10^{25}$ Mx to the solar surface
each day.

In fact, the term "horizontal magnetic elements" in this research is
not strict in a way. Because these regions selected by us are not
severely at disk center, the linear polarization also includes the
fractional information of vertical fields. Thusly, the horizontal
and vertical magnetic fields in the research are contained each
other in some sense. However, part of the 15 regions are very close
to the disk center, and the largest offset of the regions from disk
center is approximately 200 arcseconds. Considering the rather large
error possibly resulted from de-projection, we did not make the
correction of projection. The observed transverse fields are
dominantly contributed by the horizontal field component. The
properties of HIFs described in this study approximately represent
the true nature of the quiet Sun's horizontal field.

\acknowledgments

The authors are grateful to the \emph{Hinode} team for providing the
data. \emph{Hinode} is a Japanese mission developed and launched by
ISAS/JAXA, with NAOJ as a domestic partner and NASA and STFC (UK) as
international partners. The authors are also thankful to Jingyuan
Liu from Beijing Normal University. It is operated by these agencies
in cooperation with ESA and NSC (Norway). This work is supported by
the National Natural Science Foundations of China (10703007,
10873020, G10573025, 40674081, 10603008, 40890161, and 10733020),
the Chinese Academy of Sciences Project KJCX2-YW-T04, and the
National Basic Research Program of China (G2006CB806303).

\begin{table}
\begin{center}
\caption{The used data.\label{tbl-2}}
\begin{tabular}{crrrrrrrrrrr}
%\tableline\tableline
\hline
  & Date & Time Range & Location & Property \\
\hline

1 & April 7, 2007 &11:05-13:59 UT & (-5$"$, -170$"$)& ephemeral region\\
2 & April 12, 2007 &11:06-11:58 UT & (-6$"$, 0$"$)& quiet region \\
3 & April 13, 2007& 03:07-04:05 UT & (-155$"$, -169$"$)& plage region\\
4 & April 13, 2007& 04:36-05:42 UT & (-155$"$, -169$"$)& plage region \\
5 & April 13, 2007& 13:05-13:40 UT & (-98$"$, 0$"$)& quiet region\\
6 & April 13, 2007& 15:05-15:45 UT & (-80$"$, 0$"$)& quiet region \\
7 & June 1, 2007 & 14:13-15:20 UT & (-31$"$, -200$"$) & quiet region \\
8 &June 1, 2007 & 15:52-16:58 UT & (-31$"$, -200$"$) & quiet region\\
9 & June 2, 2007 & 11:48-12:40 UT & (0$"$, -200$"$) & quiet region\\
10 & June 2, 2007 & 14:51-15:57 UT & (0$"$, -200$"$) & quiet region\\
11 & June 2, 2007 & 16:29-17:35 UT & (0$"$, -200$"$) & quiet region\\
12 & June 3, 2007 & 13:50-14:56 UT & (0$"$, -200$"$) & quiet region\\
13 & June 3, 2007 & 15:28-16:34 UT & (0$"$, -200$"$) & quiet region\\
14 & June 4, 2007 & 02:58-04:04 UT & (0$"$, 200$"$) & quiet region\\
15 & November 13, 2007 & 07:06-10:00 UT & (-140$"$, -50$"$) & ephemeral region\\
\hline
16 & May 11, 2007 & 01:05-01:35 UT & (-14$"$, 0$"$) & quiet region\\

\hline
%\tableline
\end{tabular}
\end{center}
\end{table}

\begin{table}
\begin{center}
\caption{The properties of HIFs.\label{tbl-2}}
\begin{tabular}{crrrrrrrrrrr}
%\tableline\tableline
\hline
Properties & mean  & deviation & minimum & maximum \\
\hline
lifetime$_{iso}$ (min)& 4.3 & 1.9& 2.0  &  10.0 \\
lifetime$_{non-iso}$ (min) &5.9& 2.5 & 2.0& 14.0  \\
lifetime$_{all}$ (min)&  5.6& 2.4 & 2.0 & 14.0 \\
B$_{iso}$ (Mx cm$^{-2}$)& 264 & 11 & 247 & 304  \\
B$_{non-iso}$ (Mx cm$^{-2}$)&273 & 15 & 248 &330\\
B$_{all}$ (Mx cm$^{-2}$) & 271 & 15 & 247 & 330 \\
size$_{iso}$ (arcsec)&1.07&0.19&0.62&1.48\\
size$_{non-iso}$ (arcsec)&1.14&0.20&0.72&2.04\\
size$_{all}$ (arcsec)&1.13&0.20&0.62&2.04\\
\hline
 number  & & 446\\
 \hline
%\tableline
\end{tabular}
\end{center}
\end{table}

\clearpage

\begin{figure}
\resizebox{23cm}{!}{\includegraphics{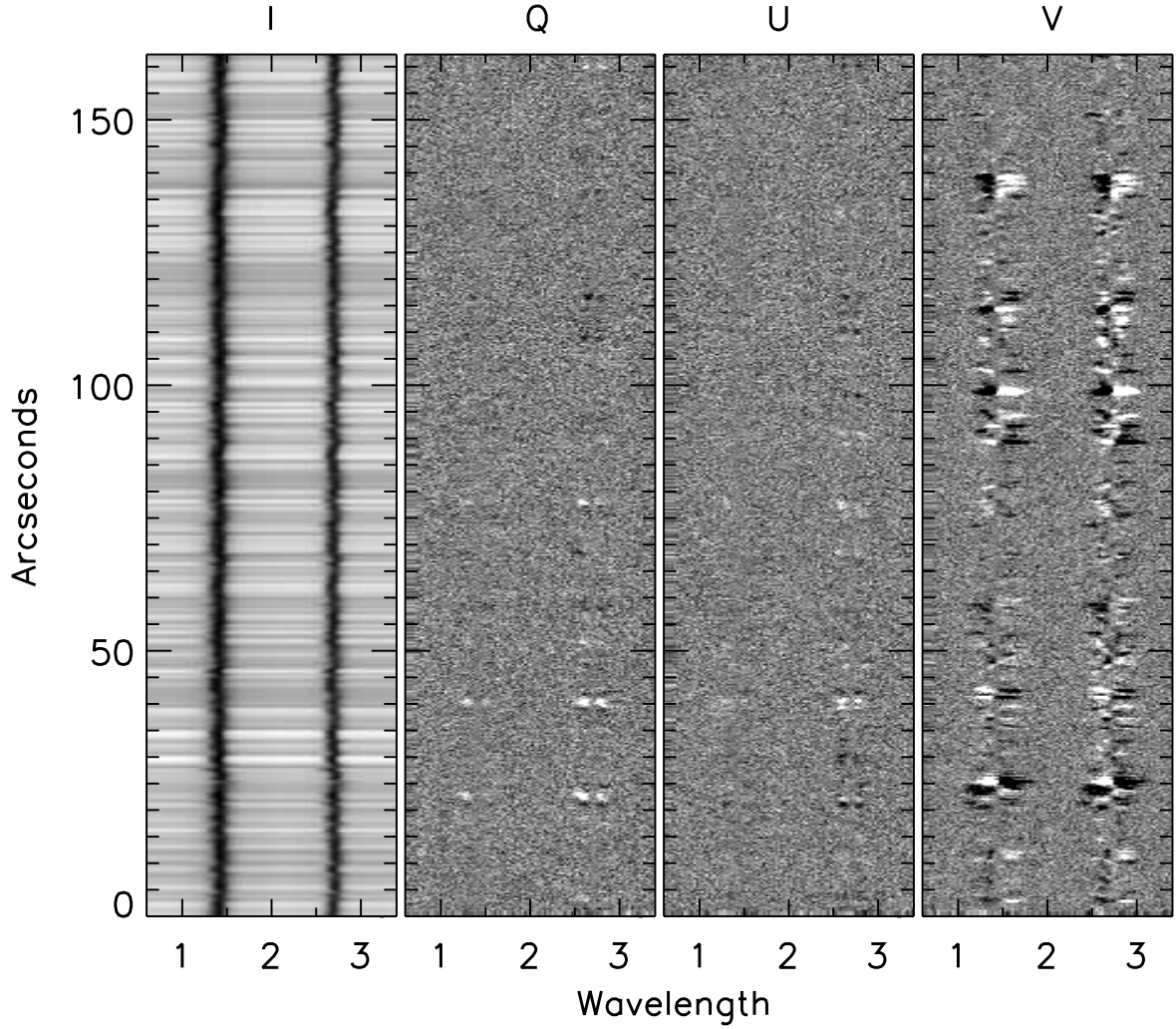}} \caption{A random
example of Stokes \emph{I}, \emph{Q}, \emph{U}, \emph{V} profiles.
These spectra were obtained near disk center at 11:08 UT on 2007
April 12 with an integration for 3.2s. The gray scale for
\emph{Q}, \emph{U}, \emph{V} saturates at $\pm$ 0.003 $I_{c}$.
There are 112 spectral samples of 2.16 pm.\label{fig1}}
\end{figure}

\begin{figure}
\resizebox{23cm}{!}{\includegraphics{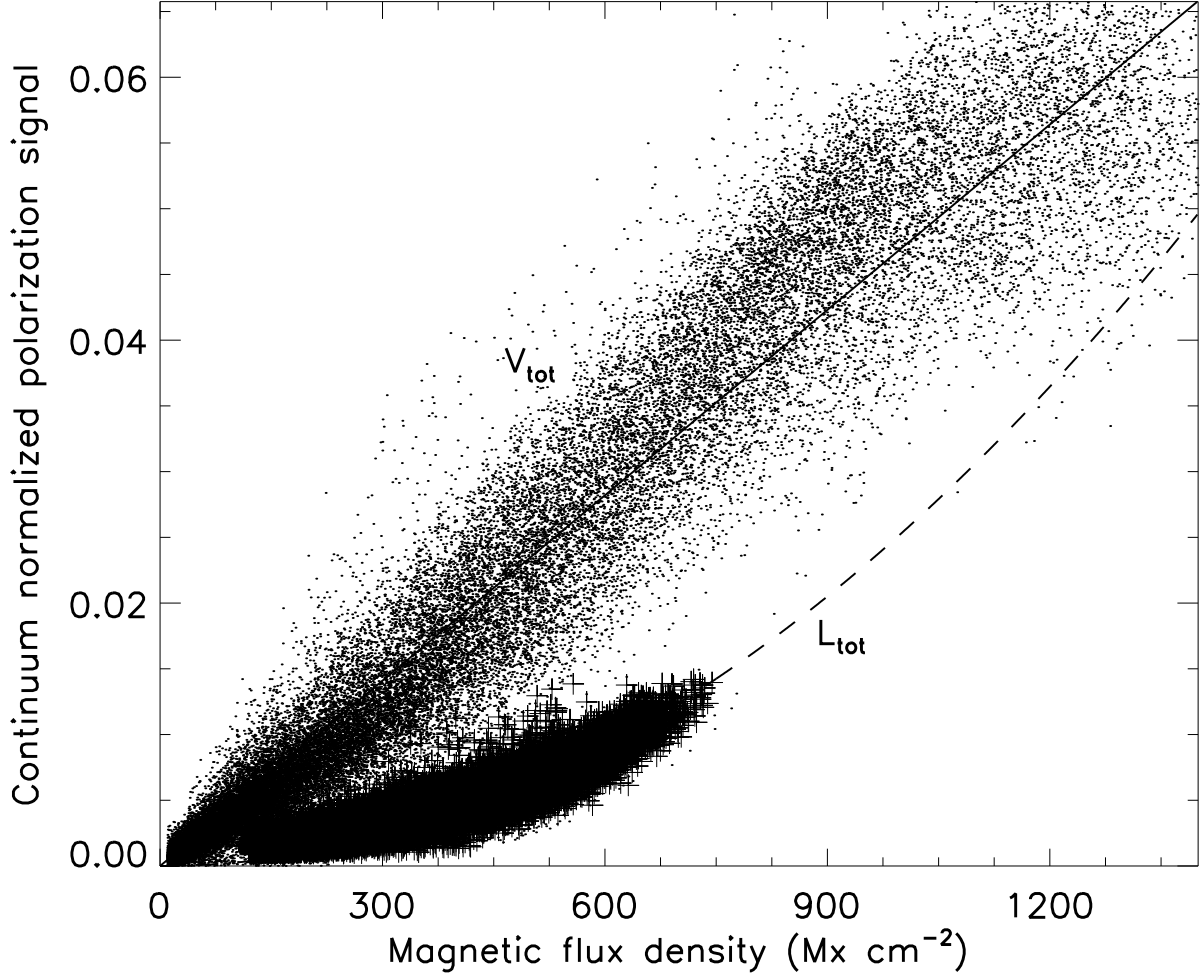}} \caption{Calibration
curves converting wavelength integral measurements of circular
polarization $V_{tot}$ (solid line) and linear polarization
$L_{tot}$ (dashed line) to magnetic flux density. The period and
plus signs mean the inversion of the stronger flux region, where
both the linear polarization and circular polarization are greater
than 3 times noise level. \label{fig2}}
\end{figure}

\begin{figure}
\resizebox{23cm}{!}{\includegraphics{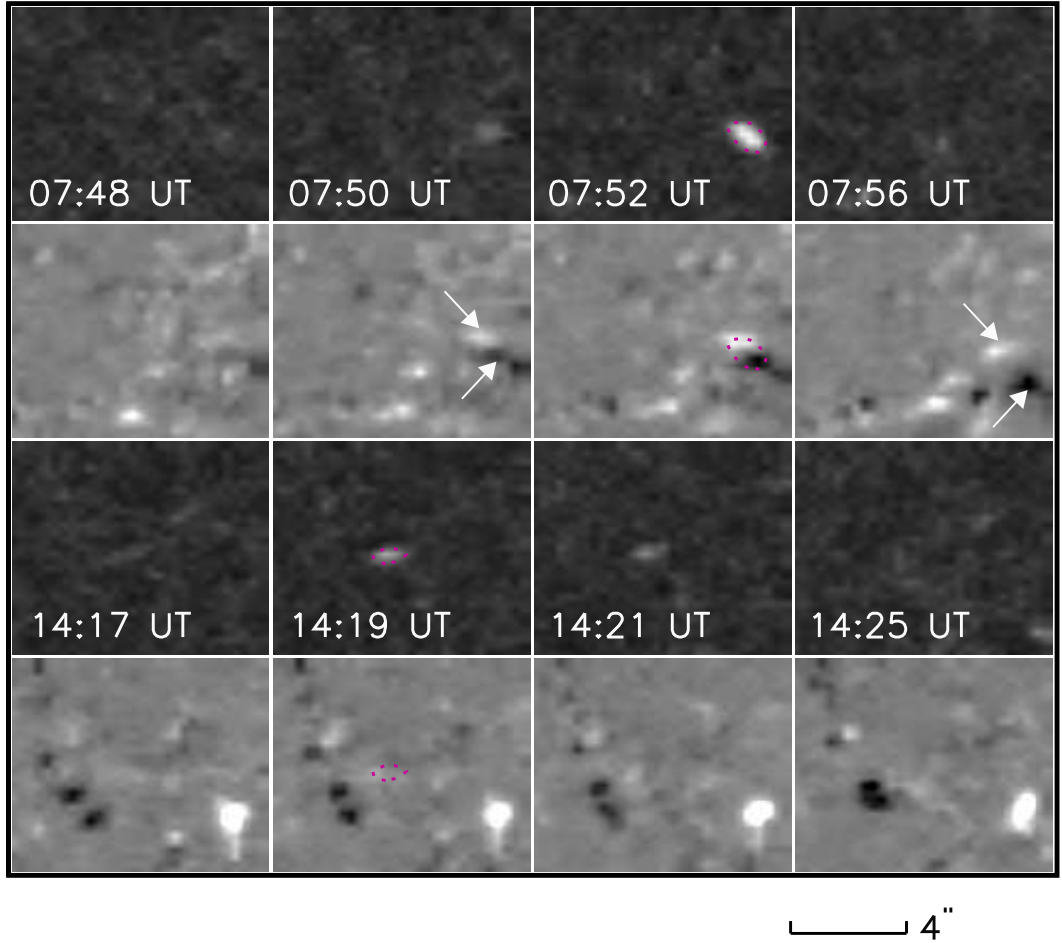}} \caption{Example of
isolated and non-isolated HIFs. The first line shows the evolution
of non-isolated HIFs observed on 2007 November 13, and the second
line is the corresponding evolution of vertical field.  The third
line describes the evolution of isolated HIFs observed on 2007
June 1, and the corresponding evolution of vertical field is shown
in the forth line. The bar in the low right corner represents the
scale of 4$''$. \label{fig3}}
\end{figure}

\begin{figure}
%\epsscale{0.50}
\resizebox{23cm}{!}{\includegraphics{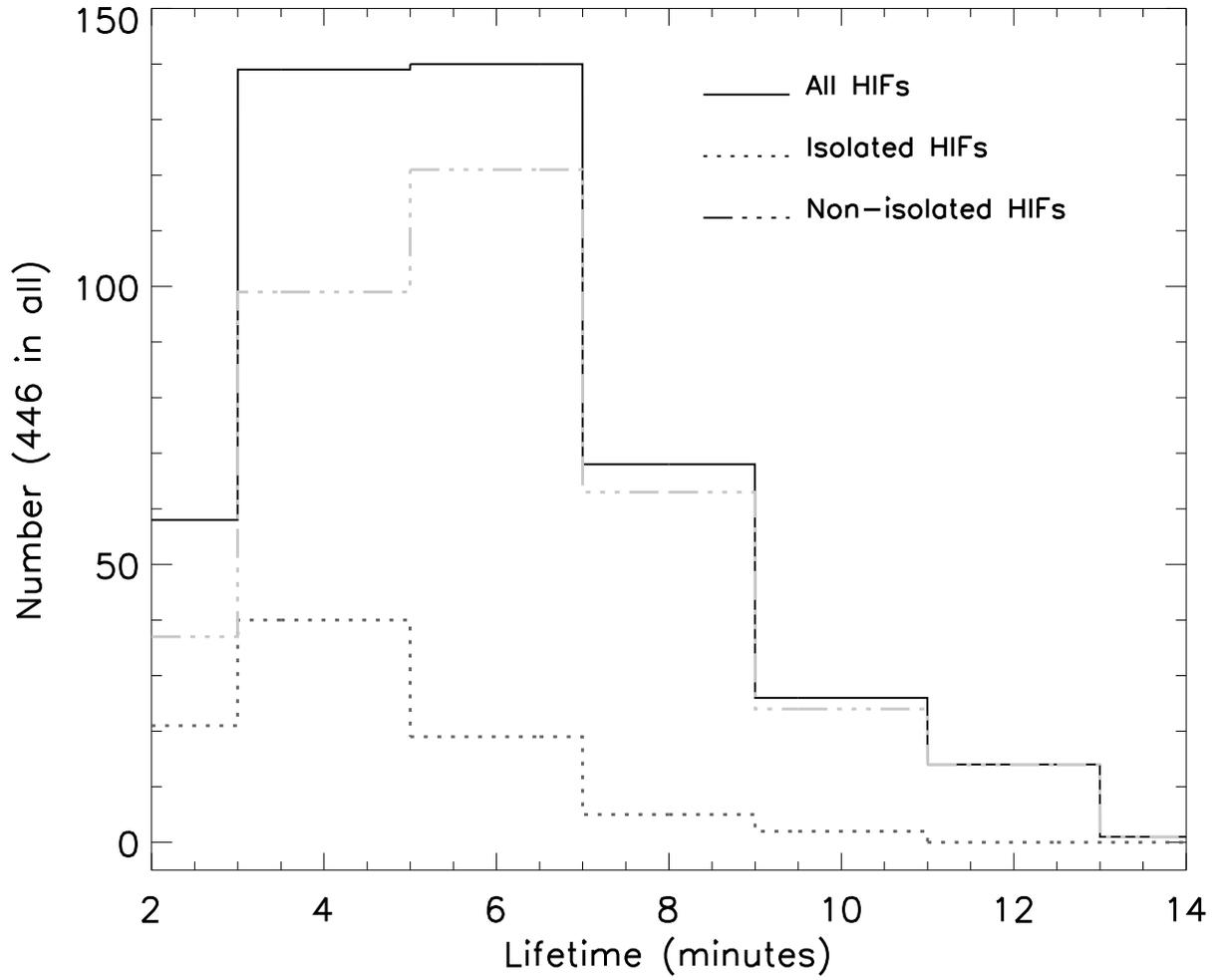}} \caption{Histogram
showing the distributions of lifetime for HIFs. The dotted line
displays the histogram for isolated HIFs, the dash-dot-dot line
for non-isolated HIFs, and the solid line for all HIFs.
\label{fig4}}
\end{figure}

\begin{figure}
\resizebox{23cm}{!}{\includegraphics{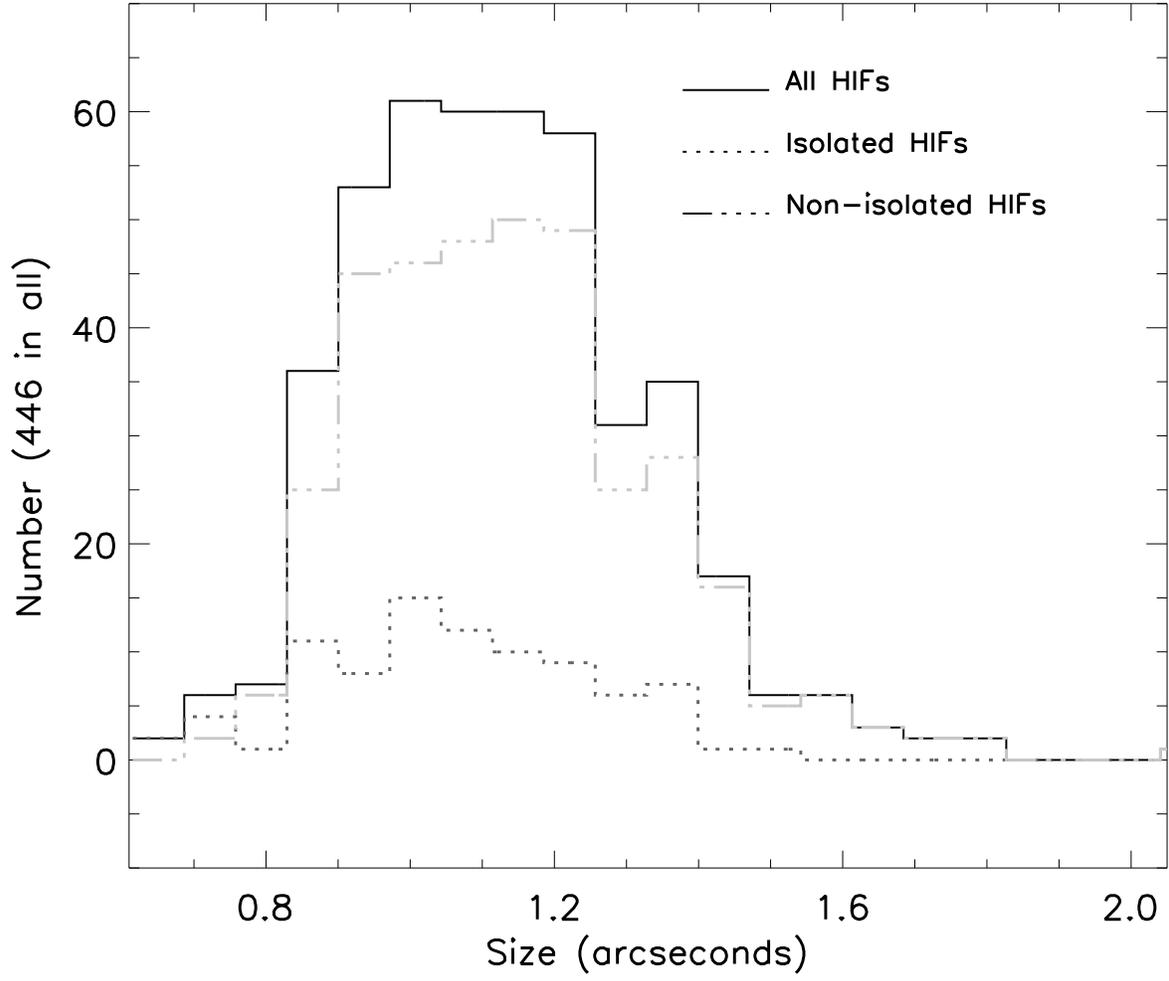}} \caption{Similar to
Fig. 4, but for the histogram of size of HIFs. \label{fig5}}
\end{figure}

\begin{figure}
%\vspace{1.2cm}
\resizebox{23cm}{!}{\includegraphics{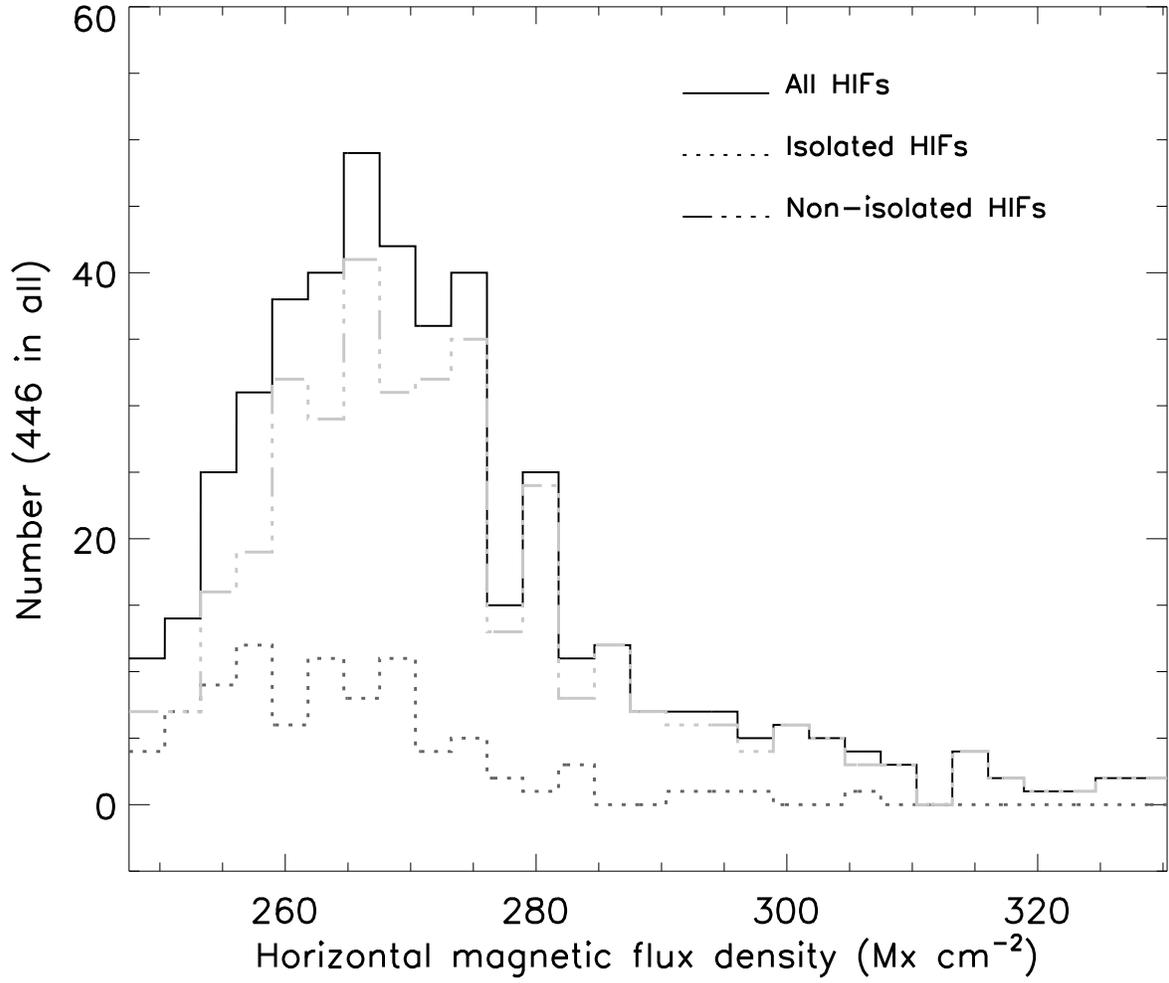}} \caption{Similar to
Fig. 4, but for the histogram of horizontal magnetic flux density
of HIFs. \label{fig6}}
\end{figure}

\begin{figure}
\resizebox{23cm}{!}{\includegraphics{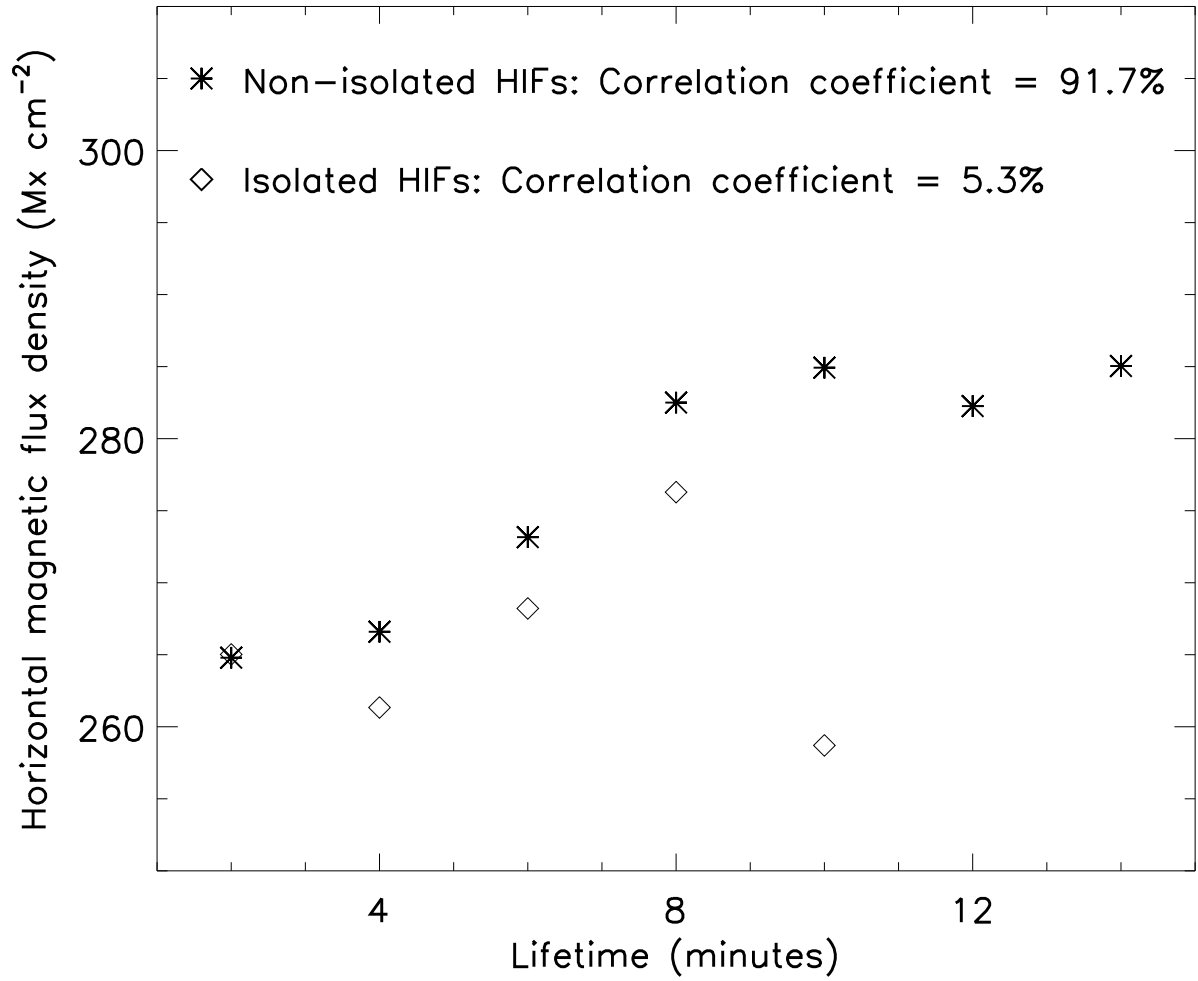}}
\caption{Relationship between lifetime and horizontal magnetic
flux density of HIFs. The asterisks mean the properties of
non-isolated HIFs, and the diamonds show the one of isolated HIFs.
\label{fig7}}
\end{figure}

\begin{figure}
\resizebox{23cm}{!}{\includegraphics{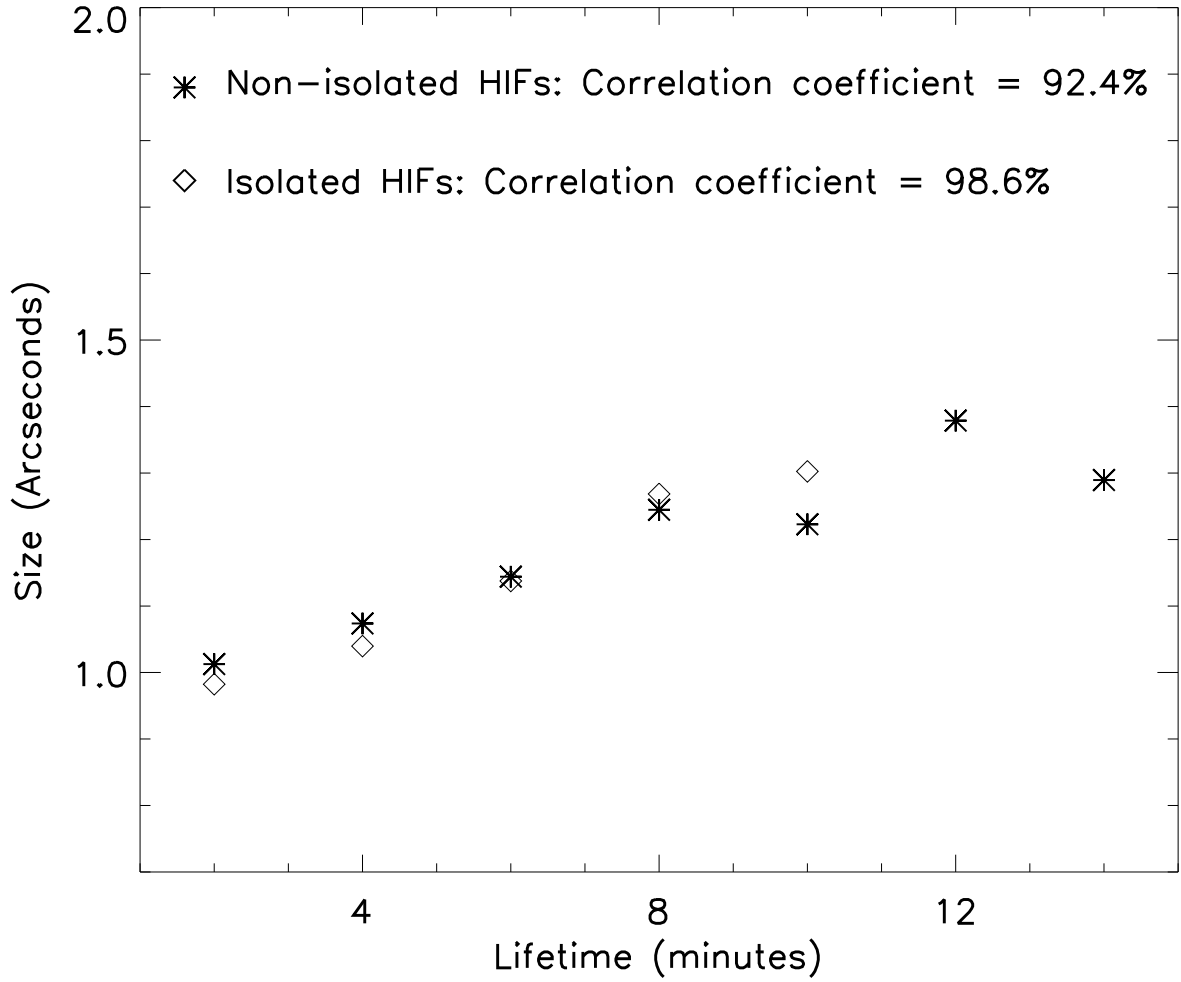}}
\caption{Relationship between lifetime and size of HIFs. The
asterisks mean the properties of non-isolated HIFs, and the
diamonds show the one of isolated HIFs. \label{fig8}}
\end{figure}

\begin{figure}
\resizebox{23cm}{!}{\includegraphics{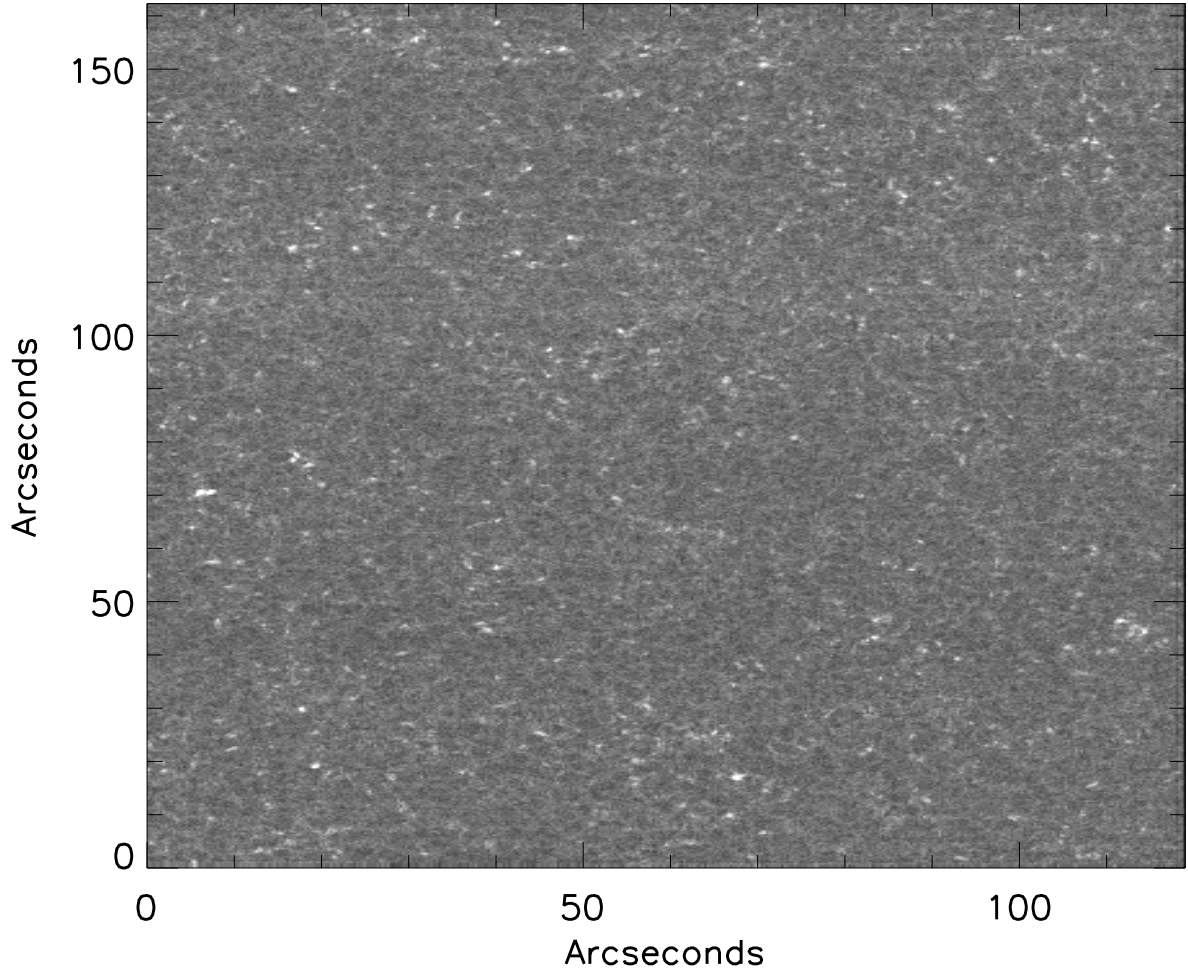}}
\caption{Distributions of horizontal magnetic flux density for
quiet region observed on 2007 May 11. The gray scale saturates at
0.0025 $I_{c}$ for linear polarization degree. \label{fig9}}
\end{figure}

\begin{figure}[htbp]
\resizebox{23cm}{!}{\includegraphics{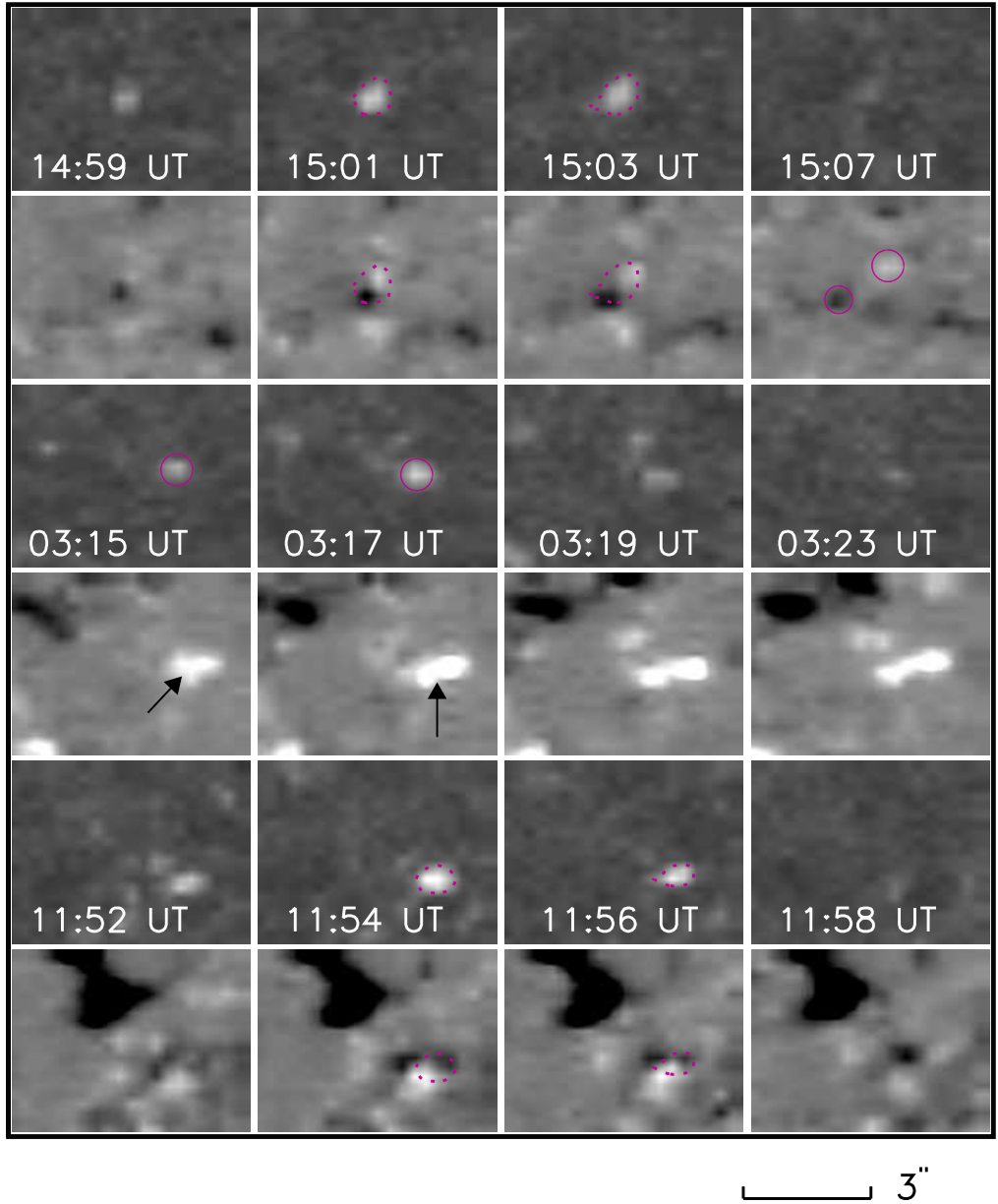}} \caption{Example
for non-isolated HIFs. The first line displays the horizontal
magnetic elements observed on 2007 June 1, and the second one
denotes the corresponding vertical magnetic elements. This is a
typical case of magnetic topology configurations for $\Omega$
loop. The polarization signals in the third and forth lines,
observed on 2007 April 13, exhibit the example of only one
footpoint in side of the region of linear polarization signal. The
magnetic elements in the fifth and sixth lines, observed on 2007
June 2, reveal the example of undistinguishable magnetic
configuration. The bar in the low right corner represents the
scale of 3$''$. \label{fig10}}
\end{figure}

\end{document}